%% file: anomaly_3_19.tex
\documentclass{elsart}
\usepackage[dvips]{graphicx}
\newcommand{\krz}{\ensuremath{K^{*0}}}
\newcommand{\krzb}{\ensuremath{\overline{K}^{*0}}}
\newcommand{\krzmndk}{\ensuremath{D^+ \rightarrow \krzb \mu^+ \nu}}

\newcommand{\kpimndk}{\ensuremath{D^+ \rightarrow K^- \pi^+ \mu^+ \nu }}

\newcommand{\gevcsq}{\ensuremath{\textrm{GeV}/c^2}}

\newcommand{\thv}{\ensuremath{\theta_\textrm{v}}}
\newcommand{\thl}{\ensuremath{\theta_\ell}}
\newcommand{\costhv}{\ensuremath{\cos\thv}}
\newcommand{\sinthv}{\ensuremath{\sin\thv}}
\newcommand{\costhl}{\ensuremath{\cos\thl}}
\newcommand{\costhlsq}{\ensuremath{\cos^2\thl}}
\newcommand{\qsq}{\ensuremath{q^2}}
\newcommand{\bw}{\ensuremath{\textrm{B}_{\krz}}}
\newcommand{\mkpi}{\ensuremath{m_{K\pi}}}
\newcommand{\amp}{\ensuremath{0.36~\exp(i \pi/4)} \ensuremath{(\textrm{GeV})^{-1}}}
\newcommand{\prd}[1]{Phys.~Rev.~D \textbf{#1}}
\newcommand{\plb}[1]{Phys.~Lett.~B \textbf{#1}}
\newcommand{\mysection}[1]{}
\newcommand{\mysubsection}[1]{}

\newcounter{saveeqn}%
\newcommand{\alphaeqn}{\setcounter{saveeqn}{\value{equation}}%
\stepcounter{saveeqn}\setcounter{equation}{0}%
\renewcommand{\theequation}
      {\mbox{\arabic{saveeqn}-\alph{equation}}}}%
\newcommand{\reseteqn}{\setcounter{equation}{\value{saveeqn}}%
\renewcommand{\theequation}{\arabic{equation}}}%

\begin{document}
\begin{frontmatter}
\title{Evidence for new interference phenomena in the decay \kpimndk}
\input{author_3_11.tex}
\nobreak
\begin{abstract}
Using a large sample of charm semileptonic decays collected by the
FOCUS photoproduction experiment at Fermilab, we present evidence for
a small, even spin $K^- \pi^+$ amplitude that interferes with the
dominant \krzb{} component in the \kpimndk{} final state. Although
this interference significantly distorts the \kpimndk{} decay angular
distributions, the new amplitude creates only a very small distortion
to the observed kaon pion mass distribution when integrated over the
other kinematic variables describing the decay. Our data can be
described by \krzb{} interference with either a constant amplitude or
broad spin zero resonance.
\end{abstract}
\end{frontmatter}
\newpage
\newpage

\mysection{Introduction}

This paper describes discrepancies between the observed decay
intensity for the decay \kpimndk{} and that expected for pure
\krzmndk{} decay. The data are overlayed with a simple model
incorporating a particular choice of interference amplitude.  A later
paper will present the results of fits for the parameters describing
the interfering amplitude along with new values of the form factors
describing \krzmndk{}.

\begin{figure}[tbph!]
 \begin{center}
  \includegraphics[width=4.0in]{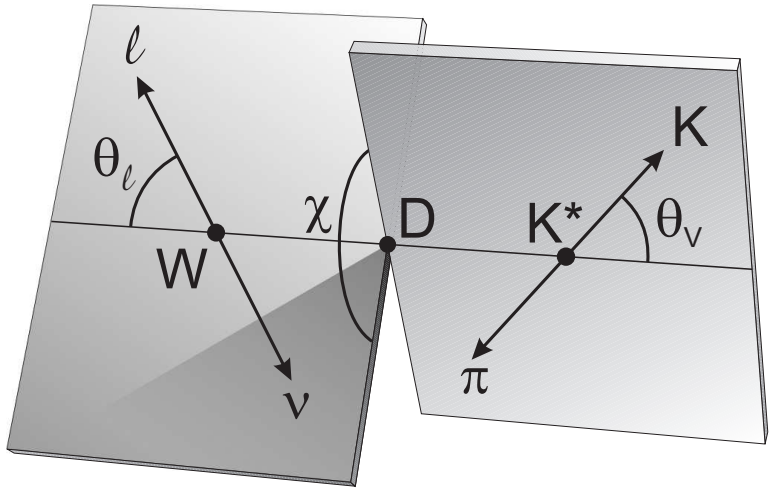}
  \caption{Definition of kinematic variables.
 \label{angles}}
 \end{center}
\end{figure}

Five kinematic variables that uniquely describe \kpimndk{} decay are
illustrated in Figure~\ref{angles}. These are the $K^- \pi^+$
invariant mass (\mkpi{}) , the square of the $\mu\nu$ mass (\qsq{}),
and three decay angles: the angle between the $\pi$ and the $D$
direction in the $K^- \pi^+$ rest frame (\thv{}), the angle between
the $\nu$ and the $D$ direction in the $\mu\nu$ rest frame (\thl{}),
and the acoplanarity angle between the two decay planes ($\chi$).  The
acoplanarity conventions for the $D^+$ and $D^-$ will be discussed 
near the end of this paper.

It has been known for many years that the \kpimndk{} final state is
strongly dominated by the \krzmndk{} channel \cite{kstardominance}.
In the process of studying the \kpimndk{} decay distribution, we found
significant discrepancies between our data and the angular
decay distributions for \krzmndk{} predicted using previously measured
~\cite{PDG} form factor ratios.  In particular, we discovered a
significant forward-backward asymmetry in \thv{} suggesting a new term
in the decay distribution that is linear in \costhv{} and a strong
function of the kaon-pion mass. In pure \krzmndk{}, the
acoplanarity-averaged decay intensity should consist of terms with
only even powers of \costhv{}.  In this paper we present an explicit
model for this effect in terms of coherent interference between a
small spin zero component of the kaon-pion system with a dominant
\krzmndk{} component.  Throughout this paper, we describe this
interference as due to a constant, s-wave amplitude described by a
single modulus and phase.  We do not exclude the possibility of a
broad spin zero resonance or even higher spin resonances with
helicity amplitudes tuned to resemble the behavior of a
single, broad s-wave resonance.

Angular momentum considerations allow us to predict the dependence of
the interference-induced, linear \costhv{} term on other decay
variables such as \qsq{}, \thl{} , and $\chi$.  We present this model
and show how well our data match these predictions. Throughout this
paper, unless explicitly stated otherwise, the charge conjugate is
also implied when a decay mode of a specific charge is stated.

\mysection{Experimental and analysis details}

The data for this paper were collected in the Wideband photoproduction
experiment FOCUS during the Fermilab 1996--1997 fixed-target run. In
FOCUS, a forward multi-particle spectrometer is used to measure the
interactions of high energy photons on a segmented BeO target. The
FOCUS detector is a large aperture, fixed-target spectrometer with
excellent vertexing and particle identification. Most of the FOCUS
experiment and analysis techniques have been described
previously~\cite{ycp}. Here, for the first time, we describe our
muon identification.
FOCUS uses two muon detector systems, the inner muon hodoscope, and
the outer muon detector.  Muons are identified by their ability to
penetrate approximately 21 interaction lengths of absorber for the
inner muon system, and 18 interaction lengths for the outer system.
For the inner system, potential muon tracks are projected through the
electromagnetic and hadronic calorimeters, and additional iron
shielding walls.  This trajectory is then matched to hits recorded in
an inner muon detector consisting of six arrays of scintillation
counters subtending approximately $\pm 45$ mrad.  The outer muon
detector consists of three views of resistive plate chambers which are
shielded by the outer electromagnetic calorimeter and the iron yoke of
the second analysis magnet.  This subtends an additional
region of roughly $\pm 140$ mrad.  For
the outer muon system, potential muon tracks are projected through the
magnet yoke and then matched to outer muon hits.


Our analysis cuts were chosen to give reasonably uniform acceptance
over the 5 kinematic decay variables, while still maintaining a
reasonably strong rejection of backgrounds.  To isolate the
\kpimndk{} topology, we required that candidate muon, pion, and kaon
tracks appeared in a secondary vertex with a confidence level
exceeding 5\%.  The muon track, when extrapolated to the shielded muon
arrays, was required to match muon hits with a confidence level
exceeding 5\%. The kaon was required to have a \v Cerenkov light
pattern more consistent with that for a kaon than that for a pion by 2
units of log likelihood, while the pion track was
required to have a light pattern favoring the pion hypothesis over
that for the kaon by 2 units~\cite{CNIM}.

To further reduce muon misidentification, an inner muon candidate was
allowed to have at most one missing hit in the 6 planes comprising our
inner muon system. In order to suppress muons from pions and kaons
decaying in our spectrometer, we required inner muon candidates to
have an energy exceeding 8 GeV.  For outer muons we required an energy
exceeding 6 GeV. Non-charm and random combinatoric backgrounds were
reduced by requiring both a detachment between the vertex containing
the $K^-\pi^+\mu^+$ and the primary production vertex of 12 standard
deviations and a minimum visible energy $(E_K+E_\pi+E_\mu)$ of 30
GeV. To suppress possible backgrounds from higher multiplicity charm
decay, we isolate the $K\pi\mu$ vertex from other tracks in the event
(not including tracks in the primary vertex) by requiring that the
maximum confidence level for another track to form a vertex with the
candidate be less than 0.1\%.

In order to allow for the missing energy of the neutrino in this
semileptonic $D^+$ decay, we required the reconstructed $K \pi \mu$
mass be less than the nominal $D^+$ mass.  Background from $D^+
\rightarrow K^- \pi^+ \pi^+$, where a pion is misidentified as a muon,
was reduced using a mass cut: we required that when these three tracks
were reconstructed as a $K \pi \pi$, their $K \pi \pi$ invariant mass
differed from the nominal $D^+$ mass by at least three standard
deviations.  In order to suppress background from $D^{*+} \rightarrow
D^0 \pi^+ \rightarrow (K^- \mu^+ \nu) \pi^+$ we required $M(K^- \mu^+
\nu \pi^+) - M(K^- \mu^+ \nu) > 0.18~\gevcsq $. The momentum of the
undetected neutrino was estimated from the $D^+$ line-of-flight as
discussed below.

We assumed that the reconstructed $D$ momentum vector points along the
direction defined by the primary and secondary vertices.  This leaves
a two-fold ambiguity on the neutrino momentum.  We use the solution
that gives the lower $D$ momentum, which, according to our Monte Carlo
studies, produced somewhat better estimates for the kinematic
variables. Due to resolution, 50\% of events were reconstructed
outside physical limits (the $p_\perp$ of the charged daughters
relative to the $D^+$ direction implied a parent mass larger than the
nominal $D^+$ mass).  These events are recovered by moving the primary
vertex to the nearest allowed solution and the kinematic variables are
recomputed.  Monte Carlo studies show that the inclusion of the
recovered events does not significantly degrade either the resolution or
the signal-to-noise ratio.

It was important to test the fidelity of the simulation with respect
to reproducing the resolution of those kinematic variables which
depend on the neutrino momentum.  To do this, we studied
fully-reconstructed $D^0 \rightarrow K^- \pi^+ \pi^+ \pi^-$ decays
where, as a test, one of the pions was reconstructed using our
line-of-flight technique. We then compared its reconstructed momentum
to its original, magnetic reconstruction in order to obtain an ``observed''
resolution function that was well matched by our simulation.

\mysection{The \kpimndk{} signal}

\begin{figure}[tbph!]
 \begin{center}
  \includegraphics[width=3.5in]{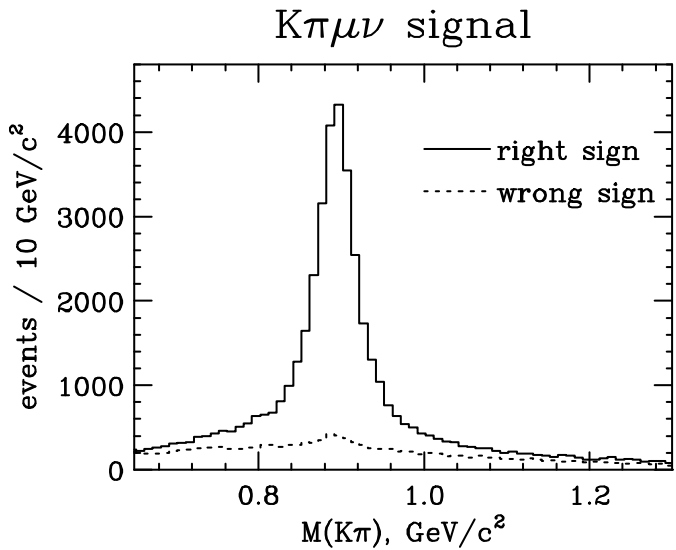}
  \caption{\kpimndk{} signal.  Right-sign and
  wrong-sign samples are shown.  The approximate wrong-sign-subtracted
  yield is $31\,254$ events.  In the mass window from 0.8--1.0 \gevcsq{}
  there is a right-sign excess of $27\,178$ events.  A Monte Carlo that
  simulates the production and decay of all known charm species
  predicts that $\approx$7\% of this excess is actually background
  from other charm decays.
\label{signal}} \end{center}
\end{figure}

Figure~\ref{signal} shows the $K^- \pi^+$ mass distribution for the
signal we obtained using the cuts described above.  A very strong
\krzb{}(896) signal is present. To assess the level of non-charm
backgrounds, we plot the ``right-sign'' (where the kaon and muon have
the opposite charge) and ``wrong-sign'' $K^- \pi^+$ mass distributions
separately. We subtract the distributions of wrong-sign
from right-sign events as a means of subtracting non-charm backgrounds
that are nearly charge symmetric.

\begin{figure}[tbph!]
 \begin{center}
  \includegraphics[width=5.5in]{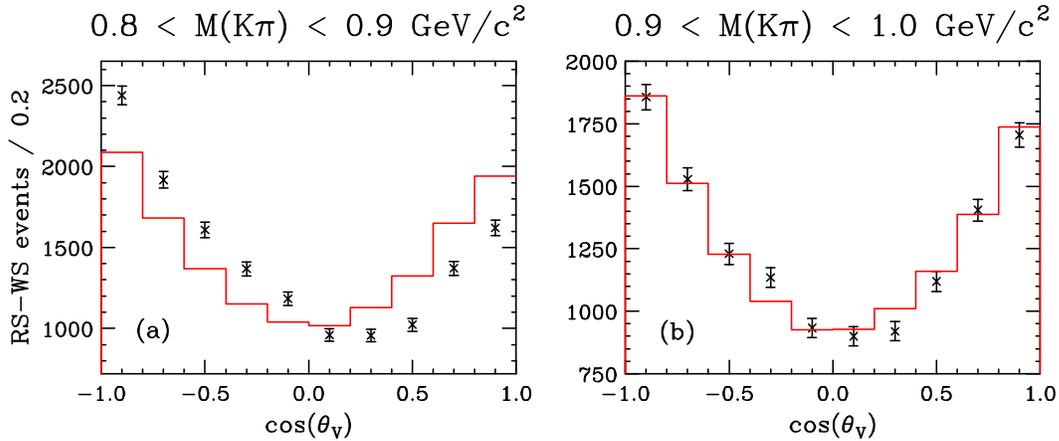}
  \caption{Event distribution in \costhv{}, split between samples
  above and below 0.9 \gevcsq{}.  The points with error bars are
  (wrong-sign subtracted) FOCUS data and the solid histogram is a Monte
  Carlo simulation, including the signal with the measured form factor
  ratios \cite{PDG} and all known charm backgrounds.  The Monte Carlo
  is normalized by area for each plot independently.}
\label{hilocosthv}
 \end{center}
\end{figure}

Figure~\ref{hilocosthv} compares the wrong-sign-subtracted \costhv{}
distribution to that predicted by our Monte Carlo which
incorporates all acceptance, and resolution effects as well as all
known charm backgrounds. We show separate distributions for events
with a $K^- \pi^+$ mass in the range $0.8 < \mkpi < 0.9 \,\gevcsq{}$ and
$0.9 < \mkpi < 1.0 \,\gevcsq{}$.

Although the events with a reconstructed mass above 0.9 \gevcsq{} are
a reasonable match to the prediction, a striking discrepancy is
apparent in the \costhv{} distribution for those events below the
pole.  As will be explained in detail later, we believe this can be
explained by the interference of a broad (or nearly constant) s-wave
amplitude with the Breit-Wigner amplitude describing the \krzb{}.  In
particular, this interference creates a term in the decay intensity
that, when averaged over acoplanarity, is linear in \costhv{} whereas
the pure \krzmndk{} process produces only even powers in \costhv{} in
this intensity. The slight forward-backward asymmetry present in the
Monte Carlo histograms of Figure~\ref{hilocosthv} primarily reflects
acceptance variation.\footnote{The known charm backgrounds tend to
have the opposite \costhv{} asymmetry to that observed in Figure
\ref{hilocosthv}. Their level can also be significantly reduced under
tighter analysis cuts that still preserve the much larger asymmetry in
the data.}

\begin{figure}[tbph!]
 \begin{center}
  \includegraphics[width=2.9in]{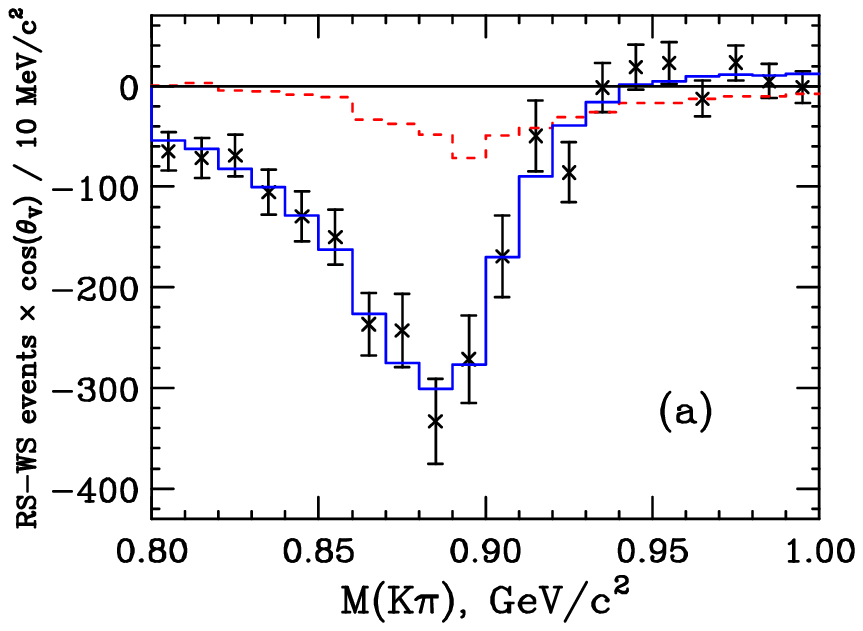}
  \includegraphics[width=2.5in]{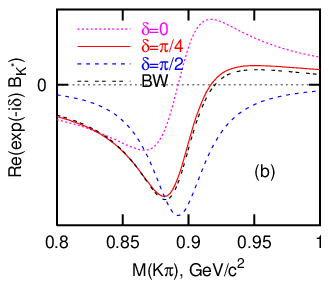}

  \caption{Asymmetry distribution in $K\pi$ invariant mass.  (a) The
  dashed line represents our Monte Carlo simulation with no
  interfering s-wave amplitude.  The experimental data are the points
  with error bars.  The solid line is the Monte Carlo with an s-wave
  amplitude of approximately 0.36 $(\textrm{GeV})^{-1}$, and a phase
  of $\frac{\pi}{4}$.  Known charm backgrounds are simulated for
  both. (b) A plot of $\Re\left(e^{-i\delta}\bw\right)$ versus \mkpi{}
  for three choices of the phase $\delta$. We also show an alternative
  modeling of the s-wave amplitude as a broad ($\Gamma =
  0.4~\gevcsq{}$) resonance with a mass of $1.1~\gevcsq{}$. We have
  put the broad, s-wave resonance in with a real phase relative to the
  \krz{} Breit-Wigner, as one might expect
given the presumed absence of final state
interactions in semileptonic decay. This resonance solution is not
  unique.  } \label{mkpiasym} \end{center}
\end{figure}

\begin{figure}[tbph!]
 \begin{center}
  \includegraphics[width=5.5in]{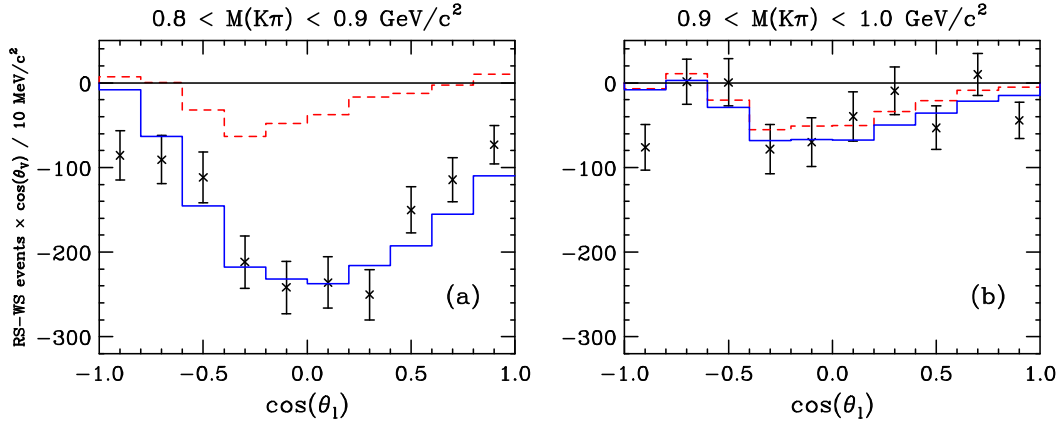}
  \caption{Asymmetry distribution in \costhl{} below (a) and above (b)
  the \krz{} pole. The dashed line represents our Monte Carlo
  simulation with no interfering s-wave amplitude.  The solid line is
  the Monte Carlo with an s-wave amplitude of approximately 0.36
  $(\textrm{GeV})^{-1}$, and a phase of $\frac{\pi}{4}$. We expect the
  acoplanarity-averaged interference to be proportional to
  $1-\costhlsq$, and to appear predominantly below the pole. Because
  the \costhv{} coefficient is negative, the weighted distribution
  appears inverted. The model with the s-wave amplitude is in good
  agreement with the data.
\label{costhlasym}} 
\end{center}
\end{figure}

\begin{figure}[tbph!]
 \begin{center}
  \includegraphics[width=3.5in]{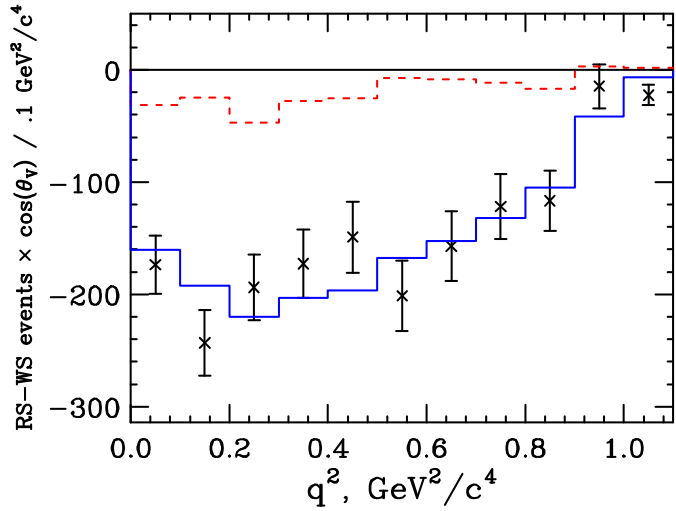}
  \caption{Asymmetry as a function of \qsq. The solid curve is our
  Monte Carlo including an interfering s-wave amplitude of \amp{}. The
  dashed curve has no interfering s-wave amplitude. Both Monte Carlo
  simulations include known charm backgrounds.  The rightmost bin also
  includes those few events where $\qsq>1.1(\gevcsq)^2$.  The model
  with the s-wave amplitude is in good agreement with the data.
\label{asymvst}}
  \end{center}
\end{figure}

We exploit the fact that all acoplanarity-averaged terms in the decay
intensity expected for pure \krzmndk{} should be proportional to even
powers of \costhv{} (see Eqn.~\ref{amp1} below).  Because our analysis
cuts give reasonably uniform acceptance over \costhv{}, we can
construct a weight designed to project out any linear \costhv{}
contribution to the decay distribution. This weight is the product of
a wrong-sign subtraction weight (+1 for right-sign and -1 for
wrong-sign) multiplied by \costhv{}. The weighted \mkpi{} distribution
is shown in Figure~\ref{mkpiasym} and compared to two Monte Carlo
simulations.  One Monte Carlo is based on pure \krzmndk{} and known
charm backgrounds (dashed histogram) while the second (solid
histogram) also includes the interfering s-wave amplitude that we
will describe later. The dashed histogram shows that the residual
effects of charm backgrounds, acceptance variation and resolution
effects produce a much smaller variation with \costhv{} than we
observe. The striking mass dependence of the linear \costhv{} term
displayed in Figure~\ref{mkpiasym} will be the principal tool we will
use to estimate the parameters of the interfering amplitude.  Figures
\ref{costhlasym}--\ref{asymvst} compare the dependence of the
\costhv{} asymmetry on two other kinematic variables (\costhl{} and
\qsq{}) to the Monte Carlo with and without the s-wave amplitude.  The
acoplanarity dependence of the interference term will be discussed in
a later section.

\mysection{Model}

It was possible to understand the forward-backward asymmetry in
\costhv{} using a simple model summarized by Eqn.~\ref{amp1}.  Using
the notation of \cite{KS}, we write the decay distribution for
\kpimndk{} in terms of the three helicity basis form factors:
$H_+~,~H_0~,~H_-$.  For simplicity, we show the decay distribution in
the limit of zero lepton mass
\begin{equation}
{d^5 \Gamma \over dm_{K \pi}~d\qsq~d\cos\thv~d\cos\thl~d\chi}
\propto K\qsq \left| \begin{array}{l}
 (1 + \cos \thl )\sin \thv e^{i\chi } \bw H_ +   \\
  - \,(1 - \cos \thl )\sin \thv e^{-i\chi } \bw H_ -   \\
  - \,2\sin \thl (\cos \thv \bw + Ae^{i\delta } )H_0  \\
 \end{array} \right|^2
\label{amp1}
\end{equation}
where $K$ is the momentum of the $K^- \pi^+$ system in the rest frame of the 
$D^+$.
The helicity basis form factors are given by:
\begin{eqnarray*}
H_\pm (\qsq) &=& (M_D+\mkpi)A_1(\qsq)\mp 2{M_D K\over M_D+m_{K\pi}}V(\qsq)\\
H_0 (\qsq) &=& {1\over 2\mkpi\sqrt{\qsq}}
\left[ (M^2_D -m^2_{K\pi}-\qsq)(M_D+\mkpi)A_1(\qsq)
-4{M^2_D K^2\over M_D+\mkpi}A_2(\qsq) \right]
\end{eqnarray*}
The vector and axial form factors are parameterized by a pole
dominance form:
\[
A_i(\qsq)={A_i(0)\over 1-\qsq/M_A^2}~~~~~~~~
V(\qsq)={V(0)\over 1-\qsq/M_V^2}
\]

where we use world average \cite{PDG} values of $R_V \equiv V(0)/
A_1(0) = 1.82 ~\rm{and}~ R_2 \equiv A_2(0)/A_1(0)= 0.78 $ and nominal
(spectroscopic) pole masses of $M_A = 2.5~\gevcsq$ and $M_V =
2.1~\gevcsq$.\footnote{Eqn.~\ref{amp1} implicitly assumes that the
\qsq{} dependence of the s-wave amplitude coupling to the virtual $W^+$
is the same as the $H_0$ form factor describing the \krzmndk{}. This
assumption is consistent with our data as illustrated in Figure
\ref{asymvst}. The modulus $A$ would then be the form factor ratio.}

The \bw{} stands for a Breit-Wigner amplitude (Eqn.~\ref{bw}) describing
the \krzb{} resonance: \footnote {We are using a p-wave Breit-Wigner
form with a width proportional to the cube of the kaon momentum in the
kaon-pion rest frame ($P^*$) over the value of this momentum when the
kaon-pion mass equals the resonant mass ($P^*_0$).}  
\begin{equation}
\bw = \frac{\sqrt{m_0 \Gamma\,} \left(\frac{P^*}{P_0^*}\right)^{(3/2)}}
	   {m_{K\pi}^2 - m_0^2 + i m_0 \Gamma 
	\left(\frac{P^*}{P_0^*}\right)^3}
\label{bw}
\end{equation}
In Eqn.~\ref{amp1}, the s-wave amplitude is modeled as a constant
(no variation with \mkpi{}) with modulus $A$ and phase
$\delta$. Angular momentum conservation restricts its contribution to
the $H_0$ piece that describes the amplitude for having the virtual
$W^+$ in a zero helicity state relative to its momentum vector in the
$D^+$ rest-frame.

Assuming this new, previously unreported amplitude is small, it will
primarily affect the decay distribution through interference with the
dominant Breit-Wigner amplitude. Expanding Eqn.~\ref{amp1}, we find
that interference between the s-wave amplitude and the \bw{} amplitude
produces the following three interference terms:
\newpage
\alphaeqn
\begin{eqnarray}
\label{inta}
\textrm{Interfere{}}
&=&8\cos\thv\sin^2\thl A
  \Re\left(e^{-i\delta}\bw\right)H_0^2  \\ 
\label{intb}
&&{}-4(1+\cos\thl)\sin\thl\sin\thv\ A
  \Re\left(e^{i(\chi-\delta)}\bw\right)H_+H_0\\ 
\label{intc}
&&{}+4(1-\cos\thl)\sin\thl\sin\thv\ A
  \Re\left(e^{-i(\chi+\delta)}\bw\right)H_-H_0
\end{eqnarray}
\reseteqn

Only the first term  (\ref{inta}),
$8\cos\thv\sin^2\thl A \Re\left(e^{-i\delta}\bw\right)H_0^2$, will be
present if one integrates over the acoplanarity variable $\chi$.
Since our acceptance is very uniform in $\chi$, we will 
primarily observe the effects of \ref{inta} when we 
average over $\chi$. We will begin by studying the
acoplanarity-averaged asymmetry distributions before turning attention
to the two acoplanarity dependent terms: Eqn.~\ref{intb} and \ref{intc}.

\mysubsection{Studies of the acoplanarity-averaged interference}

The previously mentioned, weighted \mkpi{} distribution is shown in
Figure~\ref{mkpiasym}. The shape of the \costhv{} term versus \mkpi{}
is a strong function of the interfering s-wave amplitude phase
$\delta$. Because acceptance and resolution corrections are small, the
phase can be informally determined from the \mkpi{} dependence
expected from the interference of this phase with the phase variation
expected for a Breit-Wigner, $\Re\left(e^{-i\delta}\bw\right)$.
Figure~\ref{mkpiasym}(a) demonstrates that the \costhv{} weighted
distribution in data is consistent with a constant s-wave
amplitude of the form \amp.  The magnitude of this amplitude
is the value required to match the total asymmetry in data over the
interval $0.8 < \mkpi < 0.9$ \gevcsq{}. In the discussion to follow,
the s-wave amplitude will be fixed to this value.

We next turn to a discussion of the dependence of the
acoplanarity-averaged linear \costhv{} term on other kinematic
variables.  Figure~\ref{costhlasym} compares the observed \costhv{}
weighted distribution as a function of \costhl{} to that expected in
our model using our constant s-wave amplitude of \amp{}.  We show the
distributions for \mkpi{} both above and below 0.9~\gevcsq{}.

In the context of our model, the acoplanarity-averaged, linear
\costhv{} term should be of the form: $8\cos\thv\sin^2\thl A
\Re\left(e^{-i\delta}\bw\right)H_0^2$ and is proportional to $1 -
\costhlsq{}$. This parabolic dependence is quite evident in the weighted
histogram shown in Figure~\ref{costhlasym} for those events with
$\mkpi{} < 0.9~\gevcsq{}$ where the forward-backward asymmetry is the
largest.

As a final test of the acoplanarity-averaged interference term, we
examine the \qsq{} dependence of the linear \costhv{} coefficient. In
our model, this coefficient should be proportional to $K~\qsq~H_0^2
(\qsq)$.  Figure~\ref{asymvst} compares the \costhv{} weighted
\qsq{} distribution to the data with and without the additional s-wave
amplitude of \amp{}.

\mysubsection{Studies of the acoplanarity dependence of the interference terms}

Eqn.~\ref{intb} and \ref{intc} produce an s-wave interference term
with sinusoidal variation in $\chi$ and are proportional to
\sinthv{}. In the absence of the s-wave amplitude, all acoplanarity
dependences either involve $\cos(2 \chi)$ or are odd functions of
\costhv{}.  The fact that (with the known form factors) the $H_-$ term
(Eqn.~\ref{intc}) dominates over the $H_+$ term has guided us in the
construction of a weight to study the acoplanarity dependence of the
s-wave interference. We weighted the acoplanarity distributions by
weights of the form $\sin(\chi + \delta)$ and $\cos(\chi + \delta)$
where $\delta = \pi/4$ (the phase of s-wave amplitude in our
model). We also multiplied these weights by $+1$ for right-sign events
and $-1$ for wrong-sign events to subtract away the bulk of the
non-charm background. For reasonably uniform acceptance, these weights
should average to zero for any constant or $\cos(2 \chi)$ terms or any
of the $\costhv{}\cos{\chi}$ terms present without s-wave
interference.

\begin{figure}[tbph!]
 \begin{center}
  \includegraphics[width=5.5in]{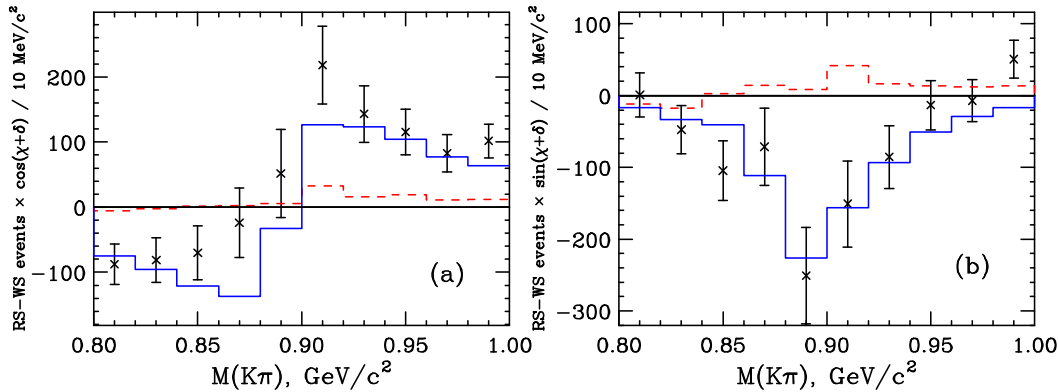}
  \caption{Test of acoplanarity interference terms. Plot (a) uses the
  $\cos(\chi+\delta)$ and is expected to be proportional to the real
  part of a \bw{} amplitude; while the plot (b) uses the $\sin(\chi +
  \delta)$ weighting and is expected to be proportional to the
  imaginary part of a \bw{} amplitude. In both figures, the solid
  histogram is our Monte Carlo including the \amp{} term; while in the
  dashed histogram there is no s-wave amplitude.
  \label{acoplanarityplot}} \end{center}
\end{figure}

Because the $H_-$ term (Eqn.~\ref{intc}) dominates over the $H_+$
term, by offsetting the phase of the weight by the phase of our s-wave
amplitude, we have arranged things such that the $\cos(\chi + \delta)$
weighted distribution should be proportional to $\Re(\bw)$ (an odd
function of \mkpi{}-$m_0$) and the $\sin(\chi + \delta)$ weighted
distribution should be proportional to $\Im(\bw)$ (an even function of
\mkpi{}-$m_0$).  These expectations are essentially borne out in the
weighted plots shown in Figure~\ref{acoplanarityplot}.

The presence of s-wave interference creates an interesting new
complication in the conventions concerning the definition of the
acoplanarity variable $\chi$.  Our conventions are guided by CP
symmetry considerations.  We define the sense of the acoplanarity
variable via a cross product expression of the form: $ (\vec P_\mu
\times \vec P_\nu) \times (\vec P_K \times \vec P_\pi) \cdot \vec P_{K
\pi}$ where all momenta vectors are in the $D^+$ rest frame.  Since
our $\chi$ convention involves five momenta vectors, we believe that
as one goes from $D^+ \rightarrow D^-$ one must change $\chi
\rightarrow -\chi$ in Eqn.~\ref{amp1}.  In the absence of the
interference, there is no need to consider the sign conventions on
$\chi$ since the decay distribution involves only cosines of $\chi$ or
$2\chi$.

\begin{figure}[tbph!]
 \begin{center} 
\includegraphics[width=5.5in]{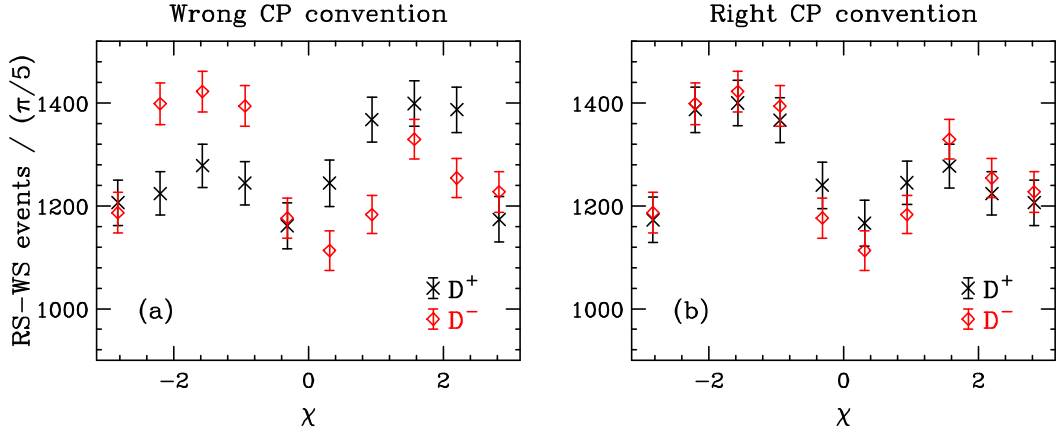}
  \caption{The wrong-sign subtracted acoplanarity distribution
  separated by charm. The ``x'' points are for the $D^+$ while the ``diamond''
  points are for the $D^-$. (a) compares the distributions without
 the required change in the $\chi$ convention as discussed above.
(b) compares the distributions with the correct $\chi$ sign convention change.
 \label{chidemo}}
  \end{center}
\end{figure}

This point is made graphically by Figure~\ref{chidemo} which compares
the observed wrong-sign subtracted acoplanarity distribution for the
$D^+$ and $D^-$.  These distributions are quite consistent once
the sign convention is properly reversed for the $D^-$ relative to the $D^+$,
indicating no evidence for CP violation in these decays. 

\mysection{Summary and Discussion}

We have presented compelling evidence for the existence of a coherent
$K^- \pi^+$ s-wave contribution to \kpimndk{}. It has been assumed in
all previous experimental analyses that this decay was strongly
dominated by the process \krzmndk{}.  The previously unobserved,
s-wave contribution is modeled as a constant amplitude of the
approximate value \amp{}.  Its strength, 0.36, is roughly 7\% of the
\krzb{} Breit-Wigner amplitude at the pole mass in the term that
couples to $H_0$ in Eqn.~\ref{amp1}.  The effect of this new
interference is very noticeable in our data and creates a
$\approx$15\% forward-backward asymmetry in the variable \costhv{} for
\krzmndk{} events with \mkpi{} below the \krzb{} pole.

\begin{figure}[tbph!]
 \begin{center}
  \includegraphics[width=4.0in]{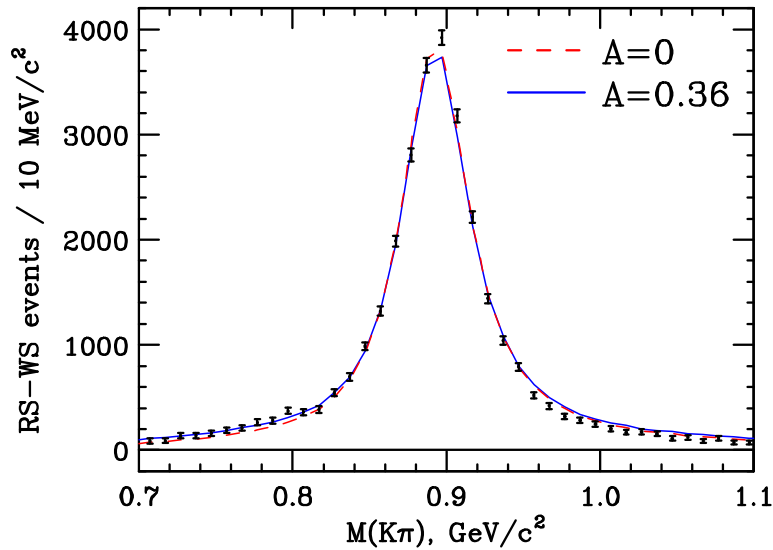}
  \caption{We show the wrong-sign subtracted \mkpi{} distribution in
  data (points with error bars) and in two Monte Carlo simulations.
  The solid simulation includes the s-wave amplitude \amp{}; while the
  dashed simulation neglects it. The known charm backgrounds are
  included in both cases.  Only a small modulation is observed
  primarily in the tails due to the inclusion of the new
  amplitude.\label{lineshape}}
\end{center}
\end{figure}

Although such an interference effect has been discussed in the
phenomenological literature \cite{phenom}, there has been no
discussion of it (to our knowledge) in the experimental literature.
How could such a large effect have gone unnoticed in the past?  We
believe one answer is that an amplitude of this strength and form
creates a very minor modulation to the \mkpi{} mass spectrum as shown
in Figure~\ref{lineshape}.  Another reason is that this effect is much
more evident when one divides the data above and below the \krzb pole
as we have done. We were unable to find evidence that this particular
split was studied in previously published data.  Finally, the FOCUS
data set has significantly more clean \kpimndk{} events than
previously published data.

\mysection{Acknowledgments}

We wish to acknowledge the assistance of the staffs of Fermi National
Accelerator Laboratory, the INFN of Italy, and the physics departments
of the collaborating institutions. This research was supported in part
by the U.~S.  National Science Foundation, the U.~S. Department of
Energy, the Italian Istituto Nazionale di Fisica Nucleare and
Ministero dell'Universit\`a e della Ricerca Scientifica e Tecnologica,
the Brazilian Conselho Nacional de Desenvolvimento Cient\'{\i}fico e
Tecnol\'ogico, CONACyT-M\'exico, the Korean Ministry of Education, and
the Korea Research Foundation.

\end{document}

%% file: author_3_11.tex
The FOCUS Collaboration%
\footnote{See \textrm{http://www-focus.fnal.gov/authors.html} for
additional author information.}
\author[ucd]{J.~M.~Link}
\author[ucd]{M.~Reyes}
\author[ucd]{P.~M.~Yager}
\author[cbpf]{J.~C.~Anjos}
\author[cbpf]{I.~Bediaga}
\author[cbpf]{C.~G\"obel}
\author[cbpf]{J.~Magnin}
\author[cbpf]{A.~Massafferri}
\author[cbpf]{J.~M.~de~Miranda}
\author[cbpf]{I.~M.~Pepe}
% ...
\address[ucd]{University of California, Davis, CA 95616}
\address[cbpf]{Centro Brasileiro de Pesquisas F\'isicas, Rio de
Janeiro, RJ, Brasil} \address[cinv]{CINVESTAV, 07000 M\'exico City,
DF, Mexico} \address[cu]{University of Colorado, Boulder, CO 80309}
\address[fnal]{Fermi National Accelerator Laboratory, Batavia, IL
60510} \address[fras]{Laboratori Nazionali di Frascati dell'INFN,
Frascati, Italy I-00044} \address[ui]{University of Illinois,
Urbana-Champaign, IL 61801} \address[iu]{Indiana University,
Bloomington, IN 47405} \address[korea]{Korea University, Seoul, Korea
136-701} \address[kp]{Kyungpook National University, Taegu, Korea
702-701} \address[milan]{INFN and University of Milano, Milano, Italy}
\address[nc]{University of North Carolina, Asheville, NC 28804}
\address[pavia]{Dipartimento di Fisica Nucleare e Teorica and INFN,
Pavia, Italy} \address[pr]{University of Puerto Rico, Mayaguez, PR
00681} \address[sc]{University of South Carolina, Columbia, SC 29208}
\address[ut]{University of Tennessee, Knoxville, TN 37996}
\address[vu]{Vanderbilt University, Nashville, TN 37235}
\address[wisc]{University of Wisconsin, Madison, WI 53706}